\documentclass[conference]{IEEEtran}

\usepackage{cite}
\usepackage{amsmath,amssymb,amsfonts}
\usepackage{graphicx}
\usepackage{textcomp}
\usepackage{xcolor}
\usepackage{booktabs}
\usepackage{array}
\usepackage{multirow}
\usepackage{hyperref}
\usepackage{cleveref}
\usepackage{balance}
\usepackage{listings}
\usepackage{subcaption}
\usepackage{pgfplots}
\usepackage[utf8]{inputenc}

\pgfplotsset{compat=1.18}

\hypersetup{
  colorlinks=true,
  linkcolor=blue,
  citecolor=blue,
  urlcolor=blue
}

\newcommand{\eg}{\textit{e.g.}}
\newcommand{\etal}{\textit{et~al.}}

\begin{document}

\title{AI-Assisted Computational Reproducibility\\on the FABRIC Testbed}

\author{
  \IEEEauthorblockN{Komal Thareja\IEEEauthorrefmark{1}, Paul Ruth\IEEEauthorrefmark{1}, Berent Aldikacti\IEEEauthorrefmark{2}, Michael Zink\IEEEauthorrefmark{2} Anirban Mandal\IEEEauthorrefmark{1}}
  \IEEEauthorblockA{\IEEEauthorrefmark{1}RENCI, University of North Carolina at Chapel Hill, NC, USA}
  \IEEEauthorblockA{\IEEEauthorrefmark{2}University of Massachusetts Amherst, MA, USA}
}

\maketitle

% ======================================================================

\begin{abstract}
Computational reproducibility remains difficult despite being central to scientific research. In this paper, we show how the international FABRIC testbed, combined with a large language model (LLM) coding agent through LoomAI, can simplify reproducing published experiments across multiple domains. We reproduced three case studies on FABRIC, covering BBR-family congestion-control evaluations, LAMMPS molecular dynamics scaling benchmarks on a CPU-only MPI cluster, and stress protein homeostasis genomics pipelines.

Rather than focusing only on matching numerical outputs, we evaluate whether the reproduced experiments support the same scientific conclusions as the original studies. The AI assistant was effective in setting up the environment, adapting code, and debugging, but struggled with the analysis stages that lacked clearly defined workflows, which required human guidance to establish execution order and data dependencies.

Across the case studies, the AI-assisted workflow reduced reproduction effort by roughly 4--6$\times$. We conclude with practical recommendations for improving AI-assisted reproducibility on research testbeds.
\end{abstract}


% ======================================================================
\section{Introduction}
\label{sec:intro}

Reproducibility is a foundational principle of science, yet computational experiments remain difficult to reproduce~\cite{baker2016reproducibility, stodden2018empirical}.
Common obstacles include incomplete environment specifications, implicit hardware assumptions, and dependency
decay~\cite{collberg2016repeatability}.  Most reproduction efforts also focus on matching numerical outputs, leaving open the more fundamental question of whether the reproduced results support the same
\emph{scientific conclusions}.

Research testbeds such as Chameleon~\cite{keahey2020chameleon}, CloudLab~\cite{cloudlab}, and FABRIC~\cite{baldin2019fabric} offer controlled, reconfigurable infrastructure on demand, but they cannot read a paper and figure out what to run. AI coding agents~\cite{chen2021codex, anthropic2024claude} can do that (read papers, generate code, debug failures), yet they have no infrastructure to execute on and cannot judge whether results are scientifically valid.  Reproduction stalls when someone must translate a paper's methodology into testbed-native code, provision the right resources, and iterate through failures.  Combining a programmable testbed (isolated slices, Software Development Kit (SDK)-defined topologies, GPU and FPGA access) with an AI coding agent (code generation, environment debugging, cross-format translation) addresses this gap, with a human researcher providing domain oversight and final validation.

We investigate this \emph{combination} using LoomAI~\cite{ruth2026loomai}, FABRIC's AI-augmented experiment
interface, to reproduce experiments from three domains. These are (1)~BBR-family congestion control (\Cref{sec:bbrv3}),
(2)~LAMMPS~\cite{lawrence2024pearc} molecular dynamics scaling (\Cref{sec:lammps}), and (3)~stress protein homeostasis genomics pipelines~\cite{aldikacti2026stress} (\Cref{sec:genomics}). An AI coding agent (Anthropic's Claude via Claude Code~\cite{anthropic2024claudecode}) drove the reproduction workflow within LoomAI, while a human researcher provided domain guidance and validated results.

The paper makes four contributions.
\begin{enumerate}
  \item[(a)] A four-phase methodology that distinguishes \emph{result-level} reproduction (do the numbers match?) from \emph{conclusion-level} reproduction (do the scientific claims still hold?), with a rubric for rating each conclusion as supported, partially supported, or not supported (\Cref{sec:methodology}).
  \item[(b)] Three cross-domain case studies in networking, molecular dynamics, and genomics, each with quantitative result comparisons and conclusion verification tables showing which original claims are independently confirmed on FABRIC (\Cref{sec:bbrv3,sec:lammps,sec:genomics}).
  \item[(c)] A quantitative analysis of AI strengths and limitations, identifying which reproducibility tasks (environment setup, code adaptation, debugging) the AI accelerated and which (domain validation, result interpretation, workflow assembly) required human expertise (\Cref{sec:ai-role}).
  %\item[(d)] A cross-domain comparison of effort, AI-generated artifacts, and reproducibility outcomes (\Cref{sec:ai-role}).
  \item[(d)] Actionable guidelines for researchers, testbed operators, and AI-tool developers to improve reproducibility outcomes (\Cref{sec:guidelines}).
\end{enumerate}

% ======================================================================
\section{Background and Related Work}
\label{sec:background}

\subsection{Computational Reproducibility}

The Association for Computing Machinery (ACM) distinguishes between \emph{repeatability} (same team, same
experimental setup), \emph{reproducibility} (different team, same experimental setup), and \emph{replicability} (different team, different experimental setup)~\cite{acm2020artifact}. We reuse each paper's original code and data but run them on different infrastructure, which changes the setup and places our work between reproducibility and replicability: rerunning GitHub artifacts on comparable Linux hosts (BBR) is closest to reproduction, while rerunning GPU benchmarks on CPU-only VMs (LAMMPS) is closer to replication.  We use \emph{reproduction} as an umbrella term and flag the distinction where it matters.  A separate axis is \emph{which} aspect is reproduced: Goodman \etal~\cite{goodman2016reproducibility} and the National Academies~\cite{nasem2019reproducibility} both separate numerical results from the inferences a paper draws.

Previous large-scale reproducibility studies have focused on single domains. Christian \etal~\cite{collberg2016repeatability} surveyed computer-systems papers, while Bhandari~Neupane \etal~\cite{bhandari2022reproducibility} examined bioinformatic pipelines.

\subsection{Research Testbeds}

\textbf{FABRIC}~\cite{baldin2019fabric} is an international programmable testbed spanning 35 sites, with 31 in the United States, three in Europe, and one in Japan.  Resources include virtual machines (VMs) with GPUs (NVIDIA RTX6000, A30, A40), Non-Volatile Memory express (NVMe) storage, 100\,Gbps links, programmable P4 switches, and Field-Programmable Gate Array (FPGA) accelerators.  Experiments are defined as \emph{slices}, isolated allocations of compute, storage, and network resources, managed through a Python SDK.

\textbf{LoomAI}~\cite{ruth2026loomai} is FABRIC's open-source, browser-based AI-augmented experiment interface.  It provides a visual topology editor, an embedded JupyterLab, integrated AI coding tools (including Claude Code, Aider, and open-weight models hosted on FABRIC GPUs), and \emph{Weaves}, which are packaged experiments that can be autonomously executed and shared via FABRIC's Artifact Manager. LoomAI was the primary environment used for all experiments in this paper.

\textbf{Chameleon}~\cite{keahey2020chameleon} offers bare-metal reconfigurable cloud infrastructure for computer-science research. \textbf{CloudLab}~\cite{cloudlab} provides similar capabilities with a focus on networking and distributed-systems experiments.

\subsection{Domain Background}

BBR~\cite{cardwell2017bbr} is a model-based TCP congestion-control family whose variants (BBRv1, BBRv2, and BBRv3~\cite{cardwell2022bbrv3}) require controlled network topologies with configurable bandwidth, delay, loss, queueing, and kernel support for evaluation. LAMMPS~\cite{thompson2022lammps} is a molecular dynamics simulator whose performance benchmarking across hardware platforms~\cite{lawrence2024pearc} tests whether scaling conclusions generalize. Modern genomics pipelines~\cite{molder2021snakemake,
ditommaso2017nextflow} process raw sequencing reads through multi-step workflows requiring matched software environments and sufficient compute resources.

\subsection{LLM Coding Assistants for Research}

LLMs trained on code~\cite{chen2021codex, anthropic2024claude, openai2023gpt4, google2024gemini} can generate, debug, and explain code across languages and scientific domains. Recent work has explored LLMs and agents for experiment design~\cite{boiko2023autonomous}, scientific prediction~\cite{sehwag2026scipredict}, and autonomous algorithm discovery~\cite{cheng2025barbarians}.  However, LLM-generated summaries overgeneralize findings nearly five times more often than human authors~\cite{peters2025generalization}, which highlights the need
for human oversight.  Adashchik \etal~\cite{adashchik2025agentic} survey agentic LLM pipelines for reproducible scientific software but note open challenges in environment replication.  Our study differs in that we pair a coding agent  with FABRIC, report quantitatively on what it could and could not do, and evaluate reproduction at the conclusion level rather than the result level.
% ======================================================================
\section{Methodology}
\label{sec:methodology}

We adopt a four-phase methodology (\Cref{fig:methodology}) applied to all three domains.  All development and execution was conducted within LoomAI~\cite{ruth2026loomai}, FABRIC's open-source AI-augmented experiment interface, using Anthropic's Claude via the Claude Code command-line interface (CLI)~\cite{anthropic2024claudecode} as the primary AI assistant.  The methodology is not specific to Claude. Any capable LLM with access to a paper's artifacts, testbed API documentation, and an execution environment could follow the same workflow (\Cref{sec:ai-role}).

\subsection{Case Study Selection}

We selected three case studies that form a deliberate gradient of reproducibility difficulty.  BBR-family networking experiments were already published on FABRIC but required integrating several papers, upstream artifacts, and kernel-specific protocol variants. LAMMPS was not originally conducted on FABRIC, but its publicly available input files and build instructions allowed a complete re-execution on FABRIC infrastructure, because the original hardware (GPUs) differs from ours (CPU-only VMs), this case is a conclusion-level replication rather than a direct reproduction and the original scaling conclusions held on the different hardware.  The genomics study was neither conducted on FABRIC nor fully deposited. Missing data and the absence of a unifying workflow definition meant that reproduction was incomplete and not all conclusions could be independently verified.

This progression, from FABRIC-native networking studies through an external HPC benchmark to an external study with incomplete deposits, tests the methodology under increasingly difficult conditions.  Each domain also exercises different FABRIC capabilities. Networking (BBR) requires programmable topology and kernel-level protocol changes, HPC (LAMMPS) requires multi-node MPI clusters with controlled interconnects, and bioinformatics (genomics) requires large-disk storage, multiple conflicting software environments, and domain-specific workflow engines.

\subsection{Phases}

\begin{figure}[t]
  \centering
  \includegraphics[width=\columnwidth]{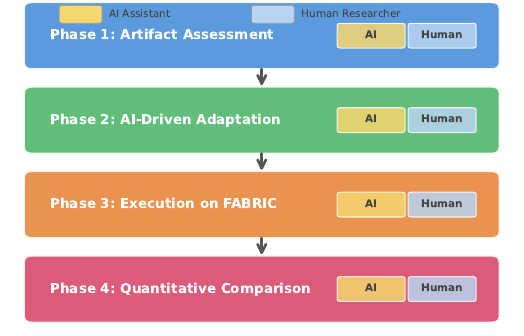}
  \caption{\small Four-phase AI-assisted reproducibility methodology.
    All phases involve both AI and human collaboration.}
  \label{fig:methodology}
\end{figure}

The workflow follows a \emph{spec-driven} development pattern (see Fig.~\ref{fig:methodology}) in which the AI first produces a design document for human review before any code is written.  Separating specification from implementation lets the researcher validate the reproduction strategy without reading code, identifies domain errors (\eg, wrong statistical test, missing data dependencies) before they propagate into artifacts, and produces an auditable record that other teams can follow independently.

In \textbf{Phase~1} (Artifact Assessment and Specification), the AI reads the target paper, public artifacts, and FABRIC-specific context (supplied via LoomAI's Retrieval-Augmented Generation (RAG) pipeline and skill plugins~\cite{ruth2026loomai}) to produce a \emph{reproduction specification}.  This structured design describes the FABRIC topology, software environment, execution steps, data dependencies, and expected outputs.  The researcher reviews and refines this specification, resolving ambiguities, correcting domain-specific details, and choosing experimental parameters, before authorizing the AI to proceed.

In \textbf{Phase~2} (AI-Driven Adaptation), the AI implements the approved specification as concrete artifacts including Jupyter notebooks that provision FABRIC slices, setup scripts, and experiment drivers.  It reuses original artifacts where available (\eg, LAMMPS input files, Snakemake workflows) and translates unavailable components from the paper's prose descriptions.  The researcher reviews generated code before execution and provides corrections.

In \textbf{Phase~3} (Execution on FABRIC), the AI executes notebooks on FABRIC nodes via Secure Shell (SSH), installs dependencies, runs experiments, and collects results, typically 5--15 tool calls per prompt.  The AI monitors for failures and iteratively fixes issues.  The researcher provides oversight when domain judgment is required, and no AI output is accepted without human review.

In \textbf{Phase~4} (Comparison and Conclusion Verification), we perform a two-level assessment.  At the \emph{result level}, we compare reproduced outputs against paper baselines using domain-appropriate metrics (\eg, speedup curves, gene counts, throughput).  At the \emph{conclusion level}, we assess whether each
scientific conclusion drawn by the original authors is \emph{supported} (reproduced evidence independently confirms the conclusion, even if numerical values differ), \emph{partially supported} (evidence direction is consistent but key data are missing or the conclusion depends on non-computational evidence), or \emph{not supported} (reproduced evidence contradicts the conclusion after accounting for expected variation).
%We define a reproduction as \emph{substantial} when $\geq$75\% of conclusion-level claims are supported or partially supported.

The verification process varied by case study.  For genomics, the original paper's authors reviewed our reproduced results and provided direct feedback on whether each conclusion was supported. For BBR and LAMMPS, conclusions were extracted from the published papers by our team. Numerical values often differ due to hardware changes even when the scientific narrative remains intact, which is why we assess conclusions rather than requiring exact numerical matches. Each case study presents a conclusion verification table.

% ======================================================================
\section{Case Study 1. BBR Network Transport}
\label{sec:bbrv3}

\subsection{Original Experiments}

This case study was our first attempt to use AI coding agents to reproduce experiments that had already run on FABRIC.  Rather than reproduce one BBRv3 paper, we selected four FABRIC-published studies of BBRv1, BBRv2, and BBRv3.  Several provided GitHub artifacts that could be rerun and updated for current tools and kernels.  The initial goal was practical. We wanted to test whether a coding agent could read the papers, locate any public artifacts, and turn them into executable FABRIC workflows with enough structure for later researchers to rerun.  \Cref{tab:bbr-paper-list} summarizes the papers, artifact sources, and outcomes.

\begin{table*}[t]
  \centering
  \caption{Selected FABRIC BBR papers, artifact sources, and reproduction outcomes.  No reproduced artifact came from the FABRIC Artifact Manager.}
  \label{tab:bbr-paper-list}
  \footnotesize
  \setlength{\tabcolsep}{2.5pt}
  \begin{tabular}{@{}>{\raggedright\arraybackslash}p{0.14\textwidth}>{\raggedright\arraybackslash}p{0.12\textwidth}>{\raggedright\arraybackslash}p{0.13\textwidth}>{\raggedright\arraybackslash}p{0.28\textwidth}>{\raggedright\arraybackslash}p{0.27\textwidth}@{}}
    \toprule
    \textbf{Paper} & \textbf{BBR Focus} & \textbf{Artifact Source} & \textbf{Original FABRIC Experiment} & \textbf{Reproduction Result Summary} \\
    \midrule
    \textbf{A. Srivastava} \etal~\cite{srivastava2023bbrdominance} & BBRv1 vs.\ CUBIC & PDF only & Shared-bottleneck dominance and Nash-equilibrium sweeps over flow mix, RTT, bottleneck rate, and buffer size. & Supported.  The 490-run rerun reproduced the BBRv1 dominance trend, with low-RTT cases still more dominant than a simple buffer-size rule predicts. \\
    \textbf{B. Gomez} \etal~\cite{gomez2023bbrv2fabric} & BBRv2 & GitHub~\cite{gomezgaona-bbr2,gomezgaona-bbr3} & Cross-site WAN throughput, RTT unfairness, queue occupancy, loss, and AQM comparisons against BBRv1, CUBIC, Reno, and H-TCP. & Partly supported.  The WAN workflow and measurement matrices ran, but \texttt{bbr2} data are BBRv3-kernel proxy results, not direct BBRv2 evidence. \\
    \textbf{C. Datta} and Fund~\cite{datta2023replicationbbr} & BBRv1, plus BBRv2 extension & GitHub~\cite{datta-imcbbrrepro} & Goodput comparison across bandwidth, RTT, and buffer grids, replicating and extending the Cao \etal BBR study~\cite{cao2019bbr}. & Supported with larger magnitude.  The 1,273-run rerun preserved the shallow-buffer BBRv1 advantage, with stronger high bandwidth-delay-product (BDP) gains than the original data. \\
    \textbf{D. Sarpkaya} \etal~\cite{sarpkaya2025bbrsharing} & BBRv1, BBRv2, BBRv3 & GitHub~\cite{sarpkaya-bbr-shared-artifact} & BBR sharing models against CUBIC and Reno using single-bottleneck and multi-sender topologies. & Partly supported.  The 609 BBRv1 runs reproduced regime-dependent model fit, but BBRv2 and BBRv3 sweeps were not completed. \\
    \bottomrule
  \end{tabular}
\end{table*}

\subsection{FABRIC Setup}

The four papers also exercised different kinds of reproduction work.  Paper~A tested whether the agent could reconstruct a complete experiment from prose and figures alone.  Paper~B tested artifact adaptation across FABRIC sites, kernels, and congestion-control variants.  Paper~C was closest to a conventional artifact rerun because the repository included the notebooks and data needed to compare against the prior Cao \etal study~\cite{cao2019bbr}.  Paper~D tested whether a newer shared-bottleneck modeling study could be rerun at scale and separated into the parts that were fully reproduced and the parts that still depended on unfinished BBRv2 and BBRv3 sweeps.

The agent was given the four PDFs, Internet access, and FABRIC context from FABlib, the portal, and public documentation.  It searched the FABRIC Artifact Manager first, but found no paper-specific artifacts for \texttt{bbr} or \texttt{congestion}.  Reusable materials came from GitHub, except for Paper~A, which was rebuilt from the PDF.  The restartable LoomAI Weave includes the PDFs, referenced artifacts, FABlib topology builders, \texttt{iperf3} outputs, parsers, and generated reports.  The human researcher selected the claims to prioritize, resolved unclear experiment parameters, and reviewed whether differences from the papers were expected variation or reproduction failures.

The four templates followed the original studies. Paper~A used a single-site 4-node line topology, Paper~B used cross-site WAN topologies, Paper~C used a 3-node line topology, and Paper~D used single-bottleneck plus multi-sender topologies.  \texttt{netem} and Linux \texttt{tc} shaped RTT, bottleneck rate, and queue size.  BDP means bandwidth-delay product, the bottleneck bandwidth multiplied by round-trip time.  CUBIC, Reno, H-TCP, and BBRv1 were available in standard images; newer BBR variants required special kernel support.  In Paper~B, data stored under \texttt{bbr2} came from a Linux~6.4 BBR-capable kernel and are treated as a BBRv3-kernel proxy, not direct BBRv2 measurements.

\subsection{Evaluation}

We evaluate at the claim level rather than requiring numerical identity, since the reruns used current FABRIC images, current package versions, and new slices.  The primary result is therefore not whether every throughput value matched, but whether the reproduced data changed the scientific interpretation of each paper.

Paper~A reproduced the BBRv1-dominance result across 490 runs.  In the 100\,Mbps, 40\,ms RTT, 2\,BDP case, one BBRv1 flow competing with nine CUBIC flows received about 37 to 39\,Mbps while each CUBIC flow received about 7\,Mbps.  The broader sweep matched the paper's pattern. Higher-BDP cases often followed the predicted majority-BBRv1 equilibrium, while low-RTT, low-BDP cases showed stronger BBRv1 dominance.  This was also the case where the agent had the least help from artifacts, since the workflow had to be reconstructed from the paper text and figures.

Paper~B produced 490 runs across throughput, RTT-unfairness, and queue-occupancy experiments.  The topology and measurement matrices were reproduced, and BBRv1 again sustained multi-Gbps WAN goodput while CUBIC stayed below roughly 1.3\,Gbps under small loss.  The main limitation is version fidelity because runs labeled BBRv2 are BBRv3-kernel proxy data.  The reproduction is therefore useful as a rerun of the FABRIC WAN workflow and a comparison among the protocols that were available, but it should not be cited as a direct confirmation of BBRv2 behavior.

Paper~C produced 1,273 CUBIC and BBRv1 runs.  With shallow 100\,KB buffers, BBRv1 outperformed CUBIC in 37 of 64 comparable scenarios and reached a 10.1$\times$ maximum goodput advantage in high-BDP cells.  With deep 10\,MB buffers, the median BBRv1/CUBIC ratio was 0.99$\times$.  The direction of the result matched the original study, but the magnitude was larger in several high-BDP cells.  This supports Datta and Fund's conclusion that BBRv1 helps most when buffers are shallow relative to path BDP, while also showing that exact goodput ratios remain sensitive to current FABRIC conditions and host configuration.  Because this paper included the strongest public artifact, it also gave the clearest comparison between an artifact-assisted rerun and a reconstructed rerun.

Paper~D produced 609 BBRv1 runs, including 504 single-loss-flow and 105 multi-flow fairness runs reported with Jain's fairness index (JFI)~\cite{jain1998fairness}.  The reproduced data support the conclusion that model accuracy depends on regime, with shallow-to-moderate buffers easier to explain than deep-buffer and many-flow settings.  The multi-flow experiments were especially useful for checking whether the modeling claims held beyond a single bottleneck flow.  This case also showed where automation stops being enough. The BBRv1 portion could be run and checked, but finishing the newer variants required kernel work and longer resource reservations.  It was the clearest indicator that an AI tool can organize and execute a reproduction, but cannot remove protocol-version requirements from the underlying system.  BBRv2 and BBRv3 sweeps were not completed, so those claims remain only partially reproduced.

% ======================================================================
\section{Case Study 2. Molecular Dynamics}
\label{sec:lammps}

\subsection{Original Experiments}

Lawrence \etal~\cite{lawrence2024pearc} compared the Kokkos and GPU acceleration packages of LAMMPS~\cite{thompson2022lammps} (Large-scale Atomic/Molecular Massively Parallel Simulator) using NVIDIA H100 and Intel Data Center GPU Max 1100 (Ponte Vecchio) accelerators on the composable ACES cluster at Texas A\&M University.  Three molecular dynamics benchmarks were used: Lennard-Jones (LJ), Embedded Atom Model (EAM), and Rhodopsin, with problem sizes of 32 million atoms (LJ/EAM) and 4 million atoms (Rhodopsin).  Strong scaling was measured by increasing the number of GPUs from 1 to 10 on a Liqid composable fabric node.

The paper makes four central conclusions. (1)~\emph{data movement strategy determines scalability}, meaning communication bandwidth, not raw compute power, is the limiting factor. (2)~The Kokkos package (all computation on GPU) outscales the GPU package (partial offload) because it minimizes host-to-GPU data transfer. (3)~The GPU package has a strong CPU core dependency and does not scale well beyond 3~GPUs on a single node. (4)~Communication dominates wall time for simple potentials (LJ, EAM),
while the more compute-intensive Rhodopsin benchmark is less affected.
Input files and build instructions are publicly
available~\cite{lawrence2024supplement}.

\subsection{FABRIC Setup}

We reproduced the same LAMMPS benchmarks as a \emph{CPU-only MPI scaling study} on FABRIC, using the exact input files from the paper's supplement~\cite{lawrence2024supplement}.  The experiment was configured on a 5-node cluster provisioned at TACC with one head node (32~cores, 128\,GB RAM) and four workers (16~cores, 64\,GB RAM each), totaling 96~cores and 384\,GB RAM connected via L2Bridge (192.168.1.0/24).  FABRIC was essential here because provisioning a composable multi-node MPI cluster with controlled L2 networking on demand is not straightforward on commodity cloud, and the SDK-defined topology ensures any FABRIC user can recreate the identical cluster from the same notebook.  The cluster ran Rocky Linux~8 with system OpenMPI and LAMMPS patch\_7Feb2024 built from source (CPU-only, MPI~+~OpenMP).

Four benchmarks were tested, namely LJ (Lennard-Jones melt, 256K atoms), EAM (bulk copper, 256K atoms), Rhodopsin (protein in membrane, 32K atoms), and SPC/E Water (water box, 36K atoms).  Both strong scaling (fixed problem size, 1--96 cores) and weak scaling (constant atoms/core, 1--96 cores) experiments were conducted with 3 repeats per
configuration.  The complete reproduction notebooks, scripts, and
results are available at~\cite{kthare10-lammps-reproducibility}.

\subsection{Evaluation}

\subsubsection{Strong Scaling}

\begin{figure}[t]
  \centering
  \includegraphics[width=\columnwidth]{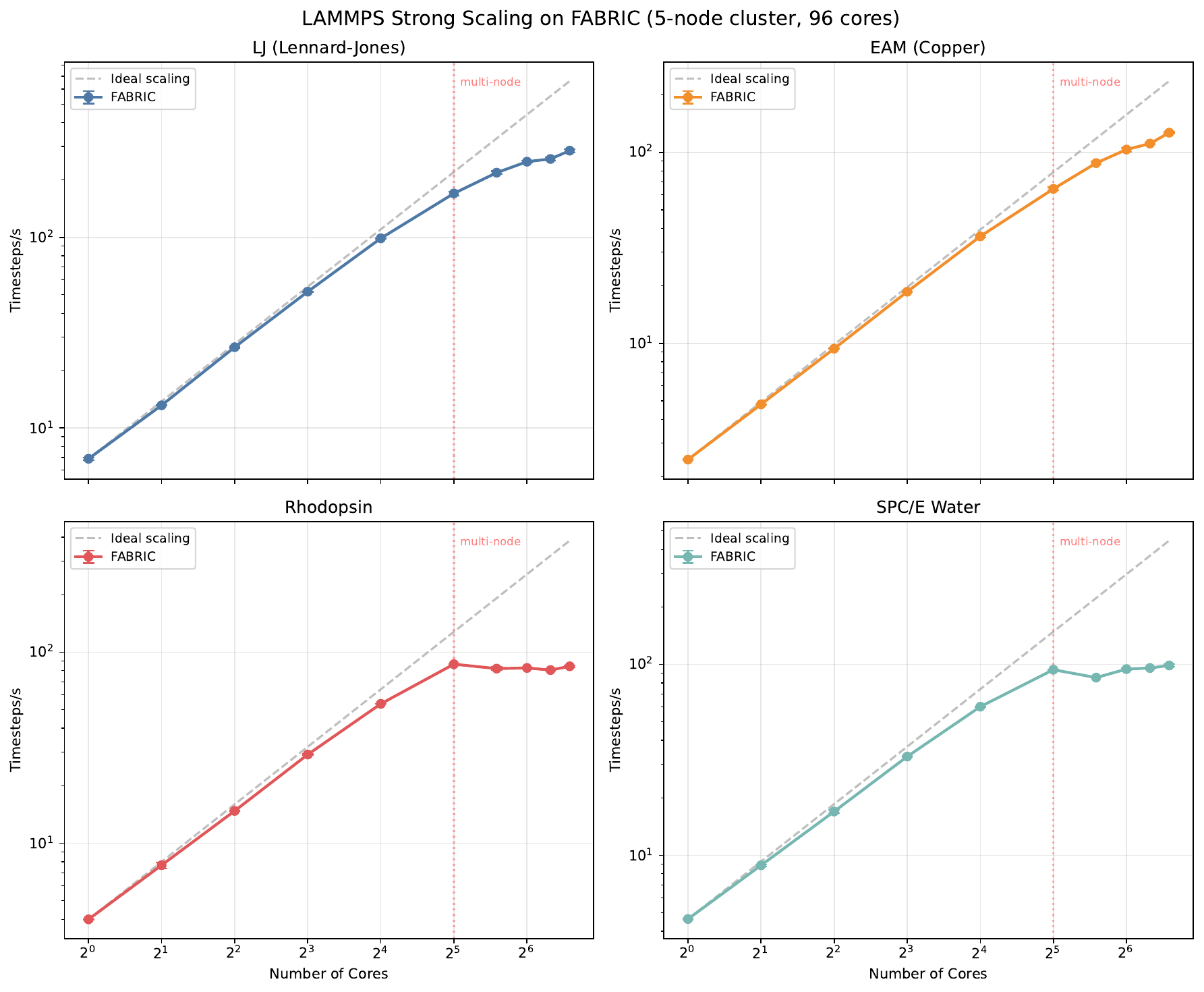}
  \caption{\small LAMMPS strong scaling on FABRIC.  Short-range potentials (LJ, EAM) track ideal scaling through 32~cores.  Long-range potentials (Rhodopsin, SPC/E) plateau at the multi-node boundary.  Dashed line = ideal linear scaling.}
  \label{fig:lammps-scaling}
\end{figure}

\Cref{fig:lammps-scaling} and \Cref{tab:lammps-strong} summarize the strong scaling results.  Short-range potentials (LJ, EAM) scaled well to 96~cores, achieving 41.4$\times$ and 51.5$\times$ speedup respectively.  Long-range potentials (Rhodopsin, SPC/E) saturated beyond 32~cores, with performance actually decreasing at the
single-to-multi-node transition (32 to 48 cores).

\begin{table}[t]
  \centering
  \caption{LAMMPS strong scaling with speedup and parallel efficiency.}
  \label{tab:lammps-strong}
  \begin{tabular}{lrrrr}
    \toprule
    \textbf{Benchmark} & \textbf{32-core} & \textbf{32-core} & \textbf{96-core} & \textbf{96-core} \\
    & \textbf{Speedup} & \textbf{Eff.} & \textbf{Speedup} & \textbf{Eff.} \\
    \midrule
    LJ        & 24.7$\times$ & 77.1\% & 41.4$\times$ & 43.1\% \\
    EAM       & 26.2$\times$ & 81.7\% & 51.5$\times$ & 53.7\% \\
    Rhodopsin & 21.6$\times$ & 67.7\% & 21.1$\times$ & 22.0\% \\
    SPC/E     & 20.2$\times$ & 63.2\% & 21.3$\times$ & 22.2\% \\
    \bottomrule
  \end{tabular}
\end{table}

\subsubsection{Communication Overhead}

Timing breakdowns reveal the scaling bottlenecks. \Cref{tab:lammps-comm} shows the fraction of wall time spent on MPI communication, Kspace (fast Fourier transform (FFT)-based long-range solver), and pair-force computation at representative core counts.

For short-range benchmarks (LJ, EAM), MPI communication grows with core count. LJ reaches 56.9\% communication at 96~cores, directly limiting further speedup.  For long-range benchmarks (Rhodopsin, SPC/E), MPI communication stays moderate (12--14\% at 96~cores), but the particle--particle particle--mesh (PPPM) Kspace solver becomes the dominant cost. Rhodopsin's Kspace fraction grows from 4.4\% to 47.2\%, and SPC/E's from 6.4\% to 53.4\%.

\begin{table}[t]
  \centering
  \caption{LAMMPS wall-time breakdown (\%) at selected core counts.}
  \label{tab:lammps-comm}
  \begin{tabular}{llrrr}
    \toprule
    \textbf{Benchmark} & \textbf{Cores} & \textbf{Comm} & \textbf{Kspace} & \textbf{Pair} \\
    \midrule
    LJ        & 1  & 0.7  & ---  & 85.0 \\
    LJ        & 32 & 24.9 & ---  & 64.0 \\
    LJ        & 96 & 56.9 & ---  & 36.4 \\
    \midrule
    Rhodopsin & 1  & 0.1  & 4.4  & 77.2 \\
    Rhodopsin & 32 & 12.7 & 14.2 & 52.1 \\
    Rhodopsin & 96 & 12.0 & 47.2 & 17.2 \\
    \midrule
    SPC/E     & 1  & 0.2  & 6.4  & 79.8 \\
    SPC/E     & 32 & 13.3 & 19.0 & 51.1 \\
    SPC/E     & 96 & 13.6 & 53.4 & 18.0 \\
    \bottomrule
  \end{tabular}
\end{table}

\subsubsection{Weak Scaling}

Weak scaling experiments held the atoms-per-core ratio approximately constant while increasing total cores from 1 to 96.  LJ retained 68.9\% efficiency at 96~cores, EAM 73.1\%, while Rhodopsin dropped to 41.2\% and SPC/E to 30.3\%, directly proportional to their communication and Kspace overhead fractions.

\subsubsection{Conclusion Verification}

\Cref{tab:lammps-conclusions} compares the paper's conclusions against our CPU-only MPI results.  Despite the entirely different hardware (CPU VMs vs.\ GPU accelerators) and communication mechanism (MPI network vs.\ host--GPU Peripheral Component Interconnect Express (PCIe) transfer), the paper's central thesis is confirmed. Data movement strategy determines scalability.  Simple potentials (LJ, EAM) with minimal computation per atom are most communication-sensitive, while the compute-intensive Rhodopsin benchmark tolerates communication overhead better.

Our experiment additionally reveals two findings not present in the original paper. First, the PPPM Kspace solver is a distinct scaling bottleneck for long-range potentials, growing to over 50\% of wall time at 96~cores independent of the MPI communication overhead. Second, the single-to-multi-node transition (32 to 48~cores) produces a sharp performance cliff for long-range potentials, with Rhodopsin and SPC/E actually losing throughput when crossing the node boundary.

\begin{table}[t]
  \centering
  \caption{Lawrence \etal~\cite{lawrence2024pearc} paper conclusions
    vs.\ FABRIC CPU-MPI reproduction.}
  \label{tab:lammps-conclusions}
  \begin{tabular}{p{4.5cm}c}
    \toprule
    \textbf{Paper Conclusion} & \textbf{Supported?} \\
    \midrule
    Data movement limits scalability & Yes \\
    Simple potentials most comm-sensitive & Yes \\
    Rhodopsin less affected by comm overhead & Yes \\
    Kokkos outscales GPU package & N/A (CPU-only) \\
    GPU package needs many CPU cores & Consistent \\
    \bottomrule
  \end{tabular}
\end{table}

The GPU-focused conclusions about communication-limited scaling generalize to CPU-only MPI on FABRIC, strengthening the original results.  The publicly available input files and build instructions made this case study the smoothest of the three.  The AI generated all 7~notebooks and 5~scripts, provisioned the multi-node cluster via the FABRIC SDK, automated MPI
job submission across core counts, parsed LAMMPS log files to extract timing breakdowns, and produced all comparison figures and tables. Human intervention was needed for choosing problem sizes (256K atoms for LJ/EAM, 32K--36K for long-range potentials), validating that the communication overhead patterns were physically meaningful, and judging whether the GPU-based scaling conclusions transferred to a CPU-only regime.

% ======================================================================
\section{Case Study 3. Genomics Pipelines}
\label{sec:genomics}

\subsection{Original Experiments}

We reproduce the computational analyses from Aldikacti \etal~\cite{aldikacti2026stress}, which studies protein homeostasis in \textit{Caulobacter crescentus} by integrating Tn-seq fitness profiling with RNA-seq expression analysis.  The pipeline spans 10~analysis stages, including Tn-seq preprocessing, ComBat-seq~\cite{zhang2020combatseq} batch correction, Generalized Linear Model (GLM) fitness classification, Model-X knockoff~\cite{candes2018knockoffs} predictor selection, Earth Mover's Distance (EMD)~\cite{rubner2000emd},
GaP-HDP~\cite{schein2021gaphdp} Bayesian clustering, RNA-seq quantification, DESeq2~\cite{love2014deseq2} differential expression, multi-omic integration, and gene network reconstruction, using code from~\cite{flahertylab-tnseq} and data from the Gene Expression Omnibus (GEO; GSE244581, GSE312471).

\subsection{FABRIC Setup}

The reproduction ran on a single FABRIC VM (24~cores, 500\,GB disk) with three isolated environments for Snakemake (Tn-seq), Nextflow (RNA-seq), and R/Python (statistical analyses).  FABRIC's large-disk VM (500\,GB) was necessary to stage the raw sequencing data, and slice-based isolation ensured the three conflicting environments (Snakemake conda, Nextflow, R) could coexist without contamination.
The entire setup is captured as a reproducible LoomAI notebook.  Raw data were downloaded from GEO/Sequence Read Archive (SRA) directly to the node.  All artifacts are available at~\cite{kthare10-stress-protein-homeostasis}.

\subsection{Evaluation}
The AI assistant generated all 27~scripts, configured the three isolated environments, and ran the 10-stage pipeline to completion on FABRIC. Stages backed by complete deposited data reproduced faithfully, while those depending on missing data could not be verified. Some conclusions were therefore untestable from incomplete artifacts, and the only tested conclusion that failed, the EMD strain ranking, itself traces to a missing input strain.  The reproduction is thus bounded more by which data were deposited than by any disagreement in the analyses we could rerun.

\begin{figure}[t]
  \centering
  \includegraphics[width=\columnwidth]{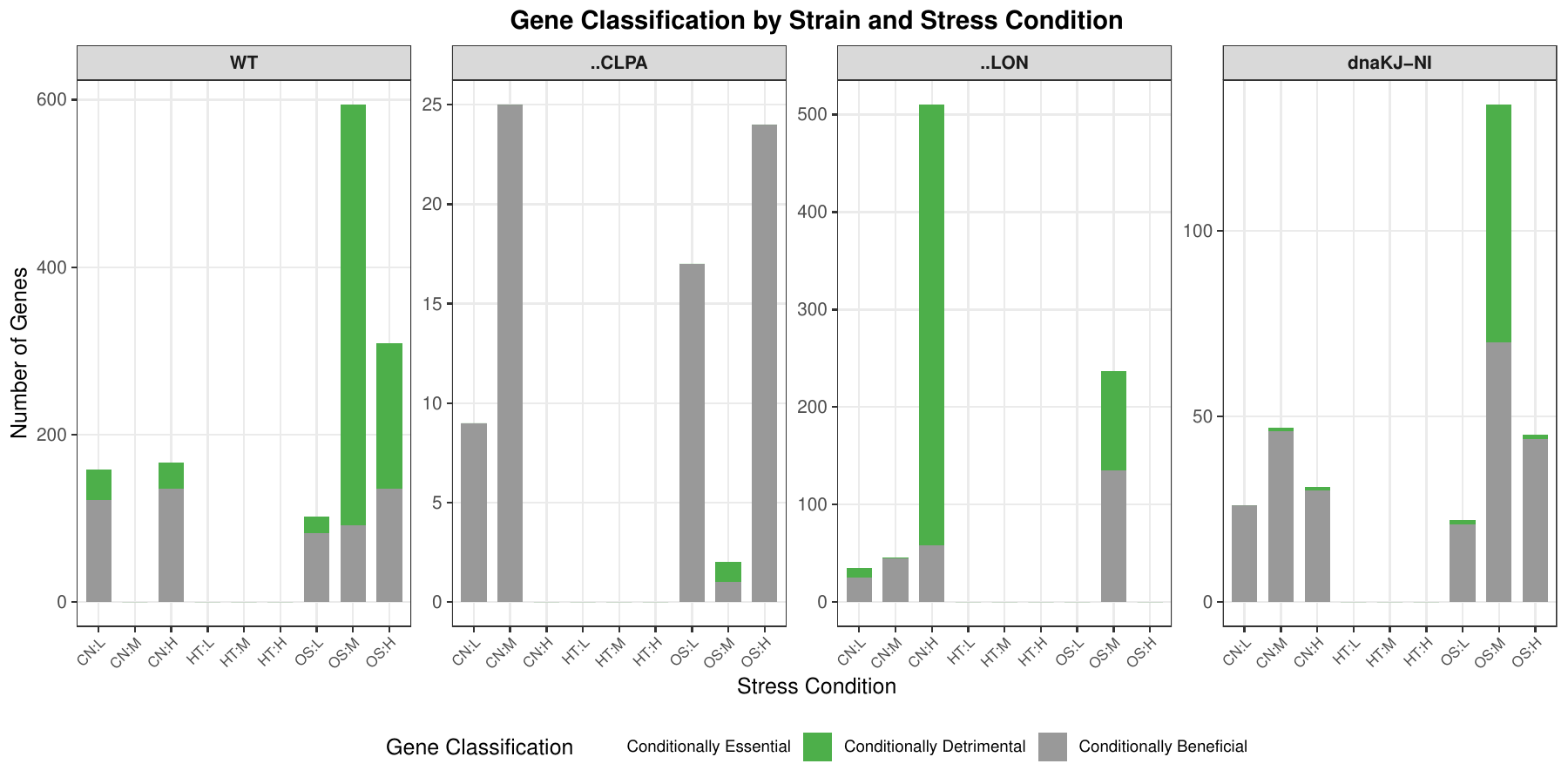}
  \caption{\small Reproduced gene classification across four strains and nine stress conditions.  Heat stress (HT:H) produces the most non-neutral genes in WT and $\Delta$lon, consistent with the paper.}
  \label{fig:genomics-classification}
\end{figure}

%The AI assistant generated all 27~scripts, configured the three isolated environments, and automated execution of the 10-stage pipeline.  All 10 analysis stages executed successfully on FABRIC, and the reproduction was assessed as \textbf{substantially reproduced} (i.e., $\geq$75\% of conclusion-level claims supported or partially supported).  
\Cref{tab:genomics-results} summarizes the quantitative comparison and \Cref{fig:genomics-classification} shows the reproduced gene classification.  Four deviations reduced quantitative confidence. These were (a)~the $\Delta$clpB Tn-seq strain was missing from the public repository (4 of 5 strains available), (b)~knockoff predictor counts deviated 16--33\% (expected for this stochastic method),
(c)~the EMD strain ranking contradicted the paper ($\Delta$lon highest vs.\ $\Delta$clpA), attributable to missing strain data, and (d)~only 14 of 18 RNA-seq samples were available.  With those caveats, the qualitative findings supported by the available data were reproduced, including the compensatory $\Delta$clpB response (2,092 vs.\ 1,592 differentially expressed (DE) genes), the ClpB $\rightarrow$ RecA $\rightarrow$ PolA pathway (35 edges), and the four-quadrant integration with ClpB correctly in Q2.

\textbf{Reproducibility barriers.}  The most significant barrier was incomplete data deposition.  Workflow-managed stages were generally easier to reproduce and required much less manual intervention. The AI assistant could only \emph{autonomously} reproduce the workflow-managed pipelines.  The remaining stages relied on ad-hoc scripts without a unifying workflow definition, and the AI could not independently determine execution order or data dependencies.  A domain biologist provided the computational schema (which inputs feed which scripts), while the AI handled mechanical adaptation (environment setup, dependency resolution, output formatting).  We conclude that, when a paper's contribution is a mathematical or biological framework rather than a software tool, the workflow is rarely organized for reproducibility, and AI assistants cannot compensate for that absence.

\begin{table}[t]
  \centering
  \caption{Genomics: paper vs.\ FABRIC results.}
  \label{tab:genomics-results}
  \begin{tabular}{lp{1.4cm}p{1.5cm}c}
    \toprule
    \textbf{Stage} & \textbf{Paper} & \textbf{Reprod.} & \textbf{Match} \\
    \midrule
    Tn-seq counts & 4{,}097 genes, 5 strains & 4{,}097 genes, 4 strains & Partial \\
    Batch correction & 694{,}280 rows & 694{,}280 rows & Yes \\
    GLM classif. & 3 categories & 1{,}113 CD, 718 CB, 160{,}684 N & Yes \\
    Knockoffs & 20--39/strain & 16--33\% dev. & Partial \\
    EMD & $\Delta$clpA highest & $\Delta$lon highest & No \\
    GaP-HDP & 121 comp. & 121 comp. & Yes \\
    RNA-seq & 18 samples & 14 samples & Partial \\
    DESeq2 & $\Delta$clpB $>$ WT & 2{,}092 vs 1{,}592 & Yes \\
    Integration & 4 quadrants & ClpB in Q2 & Yes \\
    Gene network & ClpB--RecA--PolA & 35 edges & Yes \\
    \bottomrule
  \end{tabular}
\end{table}

\subsubsection{Conclusion Verification}

\Cref{tab:aldikacti-conclusions} evaluates whether the reproduced results support the paper's eight key conclusions. Six of eight are fully supported. The ClpB pathway, compensatory $\Delta$clpB response, four-quadrant gene distribution, and functional redundancy findings all emerge from re-executed analyses. The EMD-based conclusion is not supported (strain ranking contradicts the paper, attributable to missing $\Delta$clpB data), and the stress toxicity conclusion is partially supported (computational evidence is consistent, but the full argument requires wet-lab experiments). Which conclusions can be verified depends on how complete the deposited data are. Conclusions relying on fully available RNA-seq data were all confirmed, while those depending on the missing $\Delta$clpB Tn-seq strain cannot be independently verified.

\begin{table}[t]
  \centering
  \caption{Aldikacti \etal~\cite{aldikacti2026stress} paper conclusions
    vs.\ reproduced results.}
  \label{tab:aldikacti-conclusions}
  \begin{tabular}{p{4.5cm}c}
    \toprule
    \textbf{Paper Conclusion} & \textbf{Supported?} \\
    \midrule
    Functional redundancy masks gene importance & Yes \\
    Heat stress most impactful; WT and $\Delta$lon most affected & Yes \\
    Knockoffs identify strain-specific predictors & Yes \\
    $\Delta$clpA has highest EMD divergence & No \\
    $\Delta$clpB compensatory transcriptional response & Yes \\
    Upregulation $\neq$ functional necessity & Yes \\
    ClpB in Q2; ClpB $\rightarrow$ RecA $\rightarrow$ PolA pathway & Yes \\
    Stress toxicity from specific protein loss & Partial \\
    \bottomrule
  \end{tabular}
\end{table}

% ======================================================================
\section{The Role of AI Assistants}
\label{sec:ai-role}

\Cref{tab:ai-tasks} summarizes where the AI assistant was effective and where human expertise remained necessary, based on our experience across all three case studies.

\begin{table}[t]
  \centering
  \caption{AI assistant effectiveness by task category.}
  \label{tab:ai-tasks}
  \begin{tabular}{lcc}
    \toprule
    \textbf{Task} & \textbf{AI} & \textbf{Human} \\
    & \textbf{Effective?} & \textbf{Needed?} \\
    \midrule
    Read paper \& draft reproduction spec & High & Review \\
    Parse README / install instructions & High & Minimal \\
    Generate FABRIC provisioning code & High & Review \\
    Translate / reuse original artifacts & High & Debug \\
    Generate experiment scripts from spec & Medium & Validate \\
    Debug build / dependency failures & Medium & Guide \\
    Interpret domain-specific results & Low & Essential \\
    Validate biological/physical correctness & Low & Essential \\
    Refine spec \& choose parameters & Low & Essential \\
    Assess whether results ``reproduce'' & Medium & Final call \\
    \bottomrule
  \end{tabular}
\end{table}

\subsection{AI Configuration}

All sessions used Claude Opus~4.6 via Claude Code~\cite{anthropic2024claudecode} integrated into LoomAI~\cite{ruth2026loomai}, with default API parameters (temperature~1.0, no system-prompt overrides).  Lightweight sub-tasks (file search, exploration) were delegated to Haiku~4.5.  The human--AI interaction protocol and context injection mechanism are described in \Cref{sec:methodology}.

\subsection{What AI Accelerated}

The AI was particularly effective at generating a detailed specification document that describes how to reproduce a given paper. This specification included information on FABRIC topology, the environment, execution steps, and expected outputs. Researchers could review and refine the document in collaboration with the AI before requesting code generation, which proved more efficient than repeatedly troubleshooting incomplete or faulty scripts.

For implementation, the AI typically produced correct FABRIC provisioning code, SSH configurations, and data transfer scripts on the first attempt. When installation issues arose, it analyzed log outputs and suggested fixes more quickly than manual troubleshooting, especially for cross ecosystem conflicts involving conda environments, container images, and VM packages. It also largely automated figure and table generation, reducing the manual effort required during the comparison phase.

\subsection{Where Human Expertise Remained Essential}

The AI's limitations clustered around domain judgment. In the genomics case, it initially ran analyses on reduced datasets that produced statistically meaningless results. A domain expert had to identify which analyses require the full dataset~\cite{aldikacti2026stress}. It also mischaracterized gene functional categories and chose inappropriate statistical tests until corrected by the domain biologist (\Cref{sec:genomics}).

%More broadly, the AI could compute metrics but could not judge whether a 15\% deviation indicated hardware differences or a genuine reproduction failure. While it drafted useful reproduction specifications, the researcher still had to choose problem sizes, topology parameters, and statistical methods.  The clearest limitation appeared in the genomics pipeline. Without a workflow engine defining execution order and data flow, the AI could not determine which scripts to run in what sequence.  A domain expert had to supply the computational schema before the AI could proceed (\Cref{sec:genomics}).

More broadly, the AI could compute metrics but could not judge whether a 15\% deviation indicated hardware differences or a genuine reproduction failure, and although it drafted useful specifications the researcher still chose problem sizes, topology parameters, and statistical methods. The clearest limit was the genomics pipeline: with no workflow engine defining execution order and data flow, the AI could not determine which scripts to run in what sequence, and a domain expert had to supply the computational schema before it could proceed (\Cref{sec:genomics}). The effort savings therefore do not imply hands-off automation. The AI moved the researcher's time from mechanical work toward judgment, but did not remove the need for a domain expert; without that expertise, several of these errors would likely have been accepted as valid.

\subsection{Generalizability to Other AI Assistants}

Our workflow requires long-context understanding ($>$100K tokens), code generation across languages, iterative debugging, and tool use. These capabilities are available in GPT-4, Gemini, and open-source models.  LoomAI~\cite{ruth2026loomai} supports this directly by including four open-source AI coding tools (Aider, OpenCode, Crush, Deep Agents) alongside Claude Code, all with FABRIC-specific context via RAG.  Switching backends requires a single configuration change.

\subsection{Effort and LLM Usage}

%\Cref{tab:effort-comparison} summarizes the AI-generated artifacts, effort estimates, and LLM resource consumption per case study.  The AI-assisted interaction time was approximately 12~hours for BBR (1~analysis notebook, 13~Python scripts), 6~hours for LAMMPS (7~notebooks, 5~scripts), and 10~hours for genomics (2~notebooks, 27~scripts spanning R, Python, and Bash across 3~isolated environments).  We estimate manual reproduction would require $\sim$70~person-hours for BBR, $\sim$25~person-hours for LAMMPS, and $\sim$60~person-hours for genomics, a 4--6$\times$ speedup.  These estimates are based on the authors' prior experience reproducing similar experiments without AI assistance and should be treated as order-of-magnitude approximations rather than controlled measurements.  The total API cost across all three case studies was approximately \$34~USD.

\Cref{tab:effort-comparison} summarizes the AI-generated artifacts, effort estimates, and LLM usage per case study. The AI-assisted hours are wall-clock researcher engagement and include reviewing generated code, fixing errors, and re-prompting the agent, so they are not a measure of unattended automation. Set against our manual estimates, they give a rough 4--6$\times$ reduction in effort at about \$34~USD in total API cost.

The manual baselines are the authors' recollected estimates for similar past work, produced by the same people who ran the AI-assisted experiments, so they risk overestimation bias, and we ran no human-subject comparison. The 4--6$\times$ figure indicates only an order-of-magnitude drop and not a measured speedup. A defensible number would need a controlled study with independent reproducers, which we leave to future work.

\begin{table}[t]
  \centering
  \caption{AI-assisted reproduction effort and LLM usage.}
  \label{tab:effort-comparison}
  \begin{tabular}{lrrr}
    \toprule
    \textbf{Metric} & \textbf{BBR} & \textbf{Mol.\ Dyn.} & \textbf{Genomics} \\
    \midrule
    \multicolumn{4}{l}{\textit{AI-generated artifacts}} \\
    \quad Notebooks          & 1 & 7        & 2 \\
    \quad Scripts            & 13 & 5        & 27 \\
    \quad Analysis stages    & 5 & 4        & 10 \\
    \midrule
    \multicolumn{4}{l}{\textit{Effort estimates}} \\
    \quad AI-assisted (hrs)  & $\sim$12 & $\sim$6  & $\sim$10 \\
    \quad Manual (hrs)       & $\sim$70 & $\sim$25 & $\sim$60 \\
    \quad Speedup            & $\sim$6$\times$ & $\sim$4$\times$ & $\sim$6$\times$ \\
    \midrule
    \multicolumn{4}{l}{\textit{LLM usage (Claude Opus 4.6)}} \\
    \quad Sessions           & 3 & 3 & 4 \\
    \quad User prompts       & $\sim$8 & 10 & 15 \\
    \quad Tool calls         & $\sim$90 & $\sim$80 & $\sim$117 \\
    \quad Input tokens (K)   & $\sim$280 & $\sim$500 & $\sim$870 \\
    \quad Output tokens (K)  & $\sim$50 & $\sim$40 & $\sim$63 \\
    \quad Est.\ cost (USD)   & $\sim$\$8 & $\sim$\$10 & $\sim$\$16 \\
    \bottomrule
  \end{tabular}
\end{table}

% ======================================================================
\section{Reproducibility Guidelines}
\label{sec:guidelines}

Based on our experience, we offer recommendations for researchers publishing computational results, operators of research testbeds, and developers of AI coding assistants.

\subsection{For Researchers}

The most impactful practices were as follows.
(a)~\emph{Deposit all data and code.}  Missing strains and unchecked intermediate files were the primary genomics barriers.
(b)~\emph{Automate the full pipeline.}  Whether via workflow managers, Makefiles, or end-to-end scripts, automated stages reproduced most faithfully and were most amenable to AI adaptation.
(c)~\emph{Pin every dependency} via lock files or containers.
(d)~\emph{Document hardware assumptions.}  The LAMMPS case showed conclusions generalize across platforms only when original context is clear.
(e)~\emph{Provide test datasets and expected outputs} for rapid debugging before full-scale runs.

\subsection{For Testbed Operators}

Curated base images with common scientific software would reduce the environment setup that dominates reproduction time.  Testbeds should support extended reservations or checkpoint/restart for long-running computations and publish machine-readable API documentation.

\subsection{For AI-Tool Developers}

Reproducibility demands long-context reasoning ($>$100K tokens) and iterative execution on remote infrastructure.  Domain-specific knowledge injected via RAG improved code quality in our experiments.  AI assistants should flag uncertainty when evaluating whether deviations indicate reproduction failures or expected variation.

% ======================================================================
\section{Conclusion}
\label{sec:conclusion}

We showed that FABRIC combined with an AI coding agent operating within LoomAI~\cite{ruth2026loomai} can serve as a general-purpose reproducibility platform across domains, for experiments that fit on readily provisioned testbed resources, cutting researcher effort by roughly 4--6$\times$ in our estimates at about at $\sim$\$34~USD in API cost.

The three case studies form a progression that reveals where AI-assisted reproduction succeeds and where it breaks down.  The BBR-family case study, selected from publications known to use FABRIC, was the most direct test of AI-driven reproduction on FABRIC-native networking research.  LAMMPS was not originally conducted on FABRIC and previously used GPU hardware could not be matched, yet the publicly available input files allowed a complete reproduction on a CPU-only MPI cluster, and the original scaling conclusions held on entirely different hardware. The genomics study was the most challenging because it was neither conducted on FABRIC nor fully deposited, and missing data meant that not all conclusions could be independently verified.  Across this gradient, the pattern is consistent. Numerical values differed due to hardware changes, but scientific conclusions remained intact when the artifacts were complete.  Had we evaluated success solely by numerical proximity, faithful reproductions would have been judged as failures, which is why conclusion-level assessment matters.

The genomics case also revealed that AI autonomy is bounded by workflow organization. The assistant could independently reproduce only stages managed by formal workflow engines, while ad-hoc scripts required a domain expert to provide the computational schema.  Better-structured artifacts would have helped the AI more than a more capable model.

The BBR case points to the same lesson from the opposite direction.  When papers and GitHub artifacts exposed enough experimental structure, the AI could rebuild complex FABRIC networking workflows and leave behind runnable notebooks rather than a one-time rerun.

In future work, we plan to extend the study to additional domains, build automated artifact-assessment tooling, and package completed reproductions as LoomAI Weaves that others can re-execute without repeating our effort.

\section*{Acknowledgments}
This work was supported by the National Science Foundation under grants \#1935966, \#2029295, \#2330891 and \#2513101.

% ======================================================================
% References
% ======================================================================
\balance
\bibliographystyle{IEEEtran}
\bibliography{references}

\end{document}